\documentclass[prl,aps,twocolumn,superscriptaddress,showpacs,10pt]{revtex4}
\usepackage[T1]{fontenc}
\usepackage[latin1]{inputenc}
\usepackage{amsmath}
\usepackage{amssymb}
\usepackage{dsfont}

\usepackage{color}

\makeatletter

\usepackage{hyperref}
\usepackage{amssymb}
\usepackage{graphicx}
\usepackage{epstopdf}
\usepackage{epsfig}
\usepackage{subfigure}

\makeatother

\newcommand{\ket}[1]{\left| #1 \right\rangle}

\renewcommand{\vec}[1]{\mathbf{#1}}
\newcommand{\e}{\text{e}}

\usepackage{cancel,ifthen}
\usepackage[normalem]{ulem}

\begin{document}

\title{From Anderson to anomalous localization in cold atomic gases with effective spin-orbit coupling}

\author{M. J. Edmonds}
\email{mje9@hw.ac.uk}
\affiliation{SUPA, Department of Physics, Heriot-Watt University, Edinburgh, EH14 4AS, UK}
\author{J. Otterbach}
\email{jotterbach@physics.harvard.edu}
\affiliation{Physics Department, Harvard University, Cambridge, 02139 MA, USA}
\author{R. G. Unanyan}
\affiliation{Department of Physics \& Research Center OPTIMAS, Technische Universit\"at Kaiserslautern, 67663 Kaiserslautern, Germany}
\author{M. Fleischhauer}
\affiliation{Department of Physics \& Research Center OPTIMAS, Technische Universit\"at Kaiserslautern, 67663 Kaiserslautern, Germany}
\author{M. Titov}
\affiliation{SUPA, Department of Physics, Heriot-Watt University, Edinburgh, EH14 4AS, UK}
\affiliation{Institut f\"ur Nanotechnologie, Karlsruhe Institute of Technology, 76021 Karlsruhe, Germany}
\author{P. \"Ohberg}
\affiliation{SUPA, Department of Physics, Heriot-Watt University, Edinburgh, EH14 4AS, UK}

\date{\today{}}

\begin{abstract}\noindent
We study the dynamics of a spin-orbit coupled Schr\"odinger particle with two internal degrees of freedom moving in a 1D random potential. We show that this model can be implemented using cold atoms interacting with external lasers. Numerical calculations of the density of states reveals the emergence of a Dyson-like
singularity at zero energy when the system approaches the quasi-relativistic limit of the random-mass Dirac model for large spin-orbit coupling. Simulations of the expansion of an initally localized wave-packet show a crossover from an exponential (Anderson) localization to an anomalous power-law behavior reminiscent of the zero-energy (mid-gap) state of the random-mass Dirac model. We discuss conditions under which the crossover is observable in an experiment and derive the zero-energy state, thus proving its existence under proper conditions.
\end{abstract}
\maketitle
%

Tightly confined ultracold atomic gases \cite{Bloch-2008} provide an ideal arena for studying a broad spectrum of one-dimensional (1D) quantum phenomena. Examples range from Hubbard-type models \cite{bosehubb,Morsch-2006}, pairing phenomena in Fermi gases \cite{Giorgini2008} to spin-orbit (SO) coupling for neutral atoms \cite{Lin-2011b,Campbell-2011-arXiv} and quasi-relativistic physics \cite{relbose,zwit}, generated by the additional use of external laser fields. The same experimental techniques can also be used to address  the effects of disorder in condensed matter systems. A prominent example is the demonstration of Anderson localization \cite{anderson} of a Bose-Einstein Condensate (BEC) in a 1D wave-guide \cite{matterloc1,matterloc2}. One dimensional Anderson localization is caused by destructive interference in a weak, disordered potential (referred to as diagonal disorder) and leads to exponential localization of the particles \cite{egloc,lightloc,micloc,soundloc}.
This behaviour may change in half-filled disordered metals \cite{Gogolin-1982}, random spin-Peierls and spin-ladder systems \cite{Fabrizio-1997}, or as recently
predicted in photonic systems with electromagnetically induced transparency \cite{slowlight}.
Delocalized zero-energy (mid-gap) states can emerge in these systems showing a power-law behavior for the correlations due to a Dyson singularity in the density of states
\cite{disorder_book}. Such anomalous localization originates in the chiral symmetry of the corresponding 1D Hamiltonian and can be realized in the system with off-diagonal disorder known as a random-mass Dirac model or the fluctuating gap model (FGM) \cite{powerlaw,randdirac,Millis-2000, Bartosch-1999b}.

In cold atom systems disorder is typically induced by a random potential and is thus of diagonal type. Here we show that the combination of a random potential with SO coupling induced by the motion in space dependent laser fields can lead to effective off-diagonal disorder.
Light-induced SO coupling has been shown to lead to an effective Dirac dynamics in \cite{zwit}. By investigating the density of states of the corresponding disorder model we derive conditions under which power-law localization can be observed and argue that they are indeed connected to the emergence of a Dyson singularity in the density of states (DOS). We show by simulating the time evolution of an initially localized wave packet that increasing the SO coupling drives a crossover from exponential (Anderson) localization to an anomalous power-law localization. We note that a model similar to ours has been considered in \cite{Zhu-2009}, however
only for the case of diagonal disorder and neglecting the kinetic energy part. As will be shown the latter can be neglected only under special circumstances and
drastically alters the behavior of the system leading to much richer physics.

\begin{figure}[t]
   \centering
  \epsfig{file=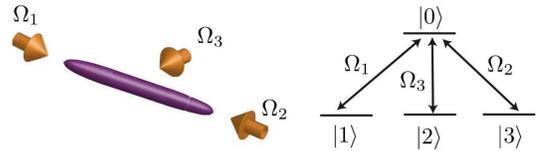,width=.4\textwidth}
   \caption{ (Color online) \textit{Left}: Experimental sketch of the system. The condensate is placed in a tight cigar-shaped trap and driven by three lasers with Rabi frequencies $\Omega_1$, $\Omega_2$ and $\Omega_3$. \textit{Right}: Internal tripod-type linkage pattern of the atoms exhibiting two dark-states without any contribution of the excited state $\ket{0}$.}
   \label{fig:TripodLinkage}
\end{figure}

We consider an ensemble of atoms exposed to three laser fields in a tripod-type linkage pattern \cite{Unanyan-1998} as depicted in Fig. (\ref{fig:TripodLinkage}). The atoms are characterized by a degenerate manifold of the three ground states $\ket{1}$, $\ket{2}$, $\ket{3}$, coupled to a common excited state $\ket{0}$ via corresponding control lasers of equal wave-number $\kappa$. We assume two of the control lasers to be counter-propagating along the $x$-axis with Rabi frequencies $\Omega_1 =\Omega \sin\theta e^{-i\kappa x}/\sqrt{2}$ and $\Omega_2 =\Omega \sin\theta e^{i\kappa x}/\sqrt{2}$, whereas the third laser propagates along the $y$-axis with $\Omega_3 =\Omega \cos\theta e^{-i\kappa y}$, and $\Omega = \sqrt{\sum_{i=1}^{3}|\Omega_i|^2}$ denotes the total Rabi frequency. Following \cite{Unanyan-1998} we can write the interaction Hamiltonian ($\hbar=1$) of the resulting tripod scheme as $
\hat{H}_{0} = -\sum_{i=1}^{3}\Big(\Omega_i|0\rangle\langle i| + h.c.\Big)$. This Hamiltonian has two dark-states,
 $|D_1\rangle  = \frac{1}{\sqrt{2}}e^{-i\kappa y}(e^{i\kappa x}|1\rangle - e^{-i\kappa x}|2\rangle)$ and $|D_2\rangle = \frac{1}{\sqrt{2}}e^{-i\kappa y}\cos\theta(e^{i\kappa x}|1\rangle + e^{-i\kappa x}|2\rangle) -\sin\theta|3\rangle$, with zero energy and decoupled from the excited state $\ket{0}$.
We express a general state of the system
in the dark state manifold according to $|\chi(\textbf{r})\rangle = \sum_{i=1}^{2}\psi_{i}(\textbf{r})|D_i(\textbf{r})\rangle$, where $\psi_1(\textbf{r})$ and $\psi_2(\textbf{r})$ are the wave functions corresponding to the two degenerate dark states. An effective equation for the center-of-mass amplitudes $\psi_i(\textbf{r})$ is then given by \cite{relbose}
\begin{equation}
i\frac{\partial \pmb\Psi}{\partial t} = \left(\frac{1}{2m}(\textbf{p} - \textbf{A})^2 + \textbf{V} + \boldsymbol\Phi\right)\pmb\Psi,
\label{eq:EffectiveGaugeFieldHamiltonian}
\end{equation}
where $\textbf{p}$ is the momentum operator, $m$ the atomic mass, and $\pmb\Psi({\bf r})=(\psi_1({\bf r}),\psi_2({\bf r}))^T$ a two-component vector.
The gauge potential $\textbf{A}$, known as the Mead-Berry connection \cite{berry,mead}, arises from the
coordinate-dependence of the dark-states and is given by $\textbf{A}_{k,n} = i\langle D_k(\textbf{r})|\nabla|D_n(\textbf{r})\rangle$. The external potential
 has matrix elements $\textbf{V}_{k,n} = \langle D_k(\textbf{r})|\hat{V}|D_n(\textbf{r})\rangle$, where the potential in the bare basis
was assumed to be diagonal, {\it i.e.}, $\hat{V}=\sum_{i=1}^{3}V_{i}(\textbf{r})|i\rangle\langle i|$. The scalar potential
reads $\boldsymbol\Phi_{k,n}=\sum_{l=1}^2\vec{A}_{k,l}\vec{A}_{l,n} /2m$.

\begin{figure}[t]
   \centering
  \epsfig{file=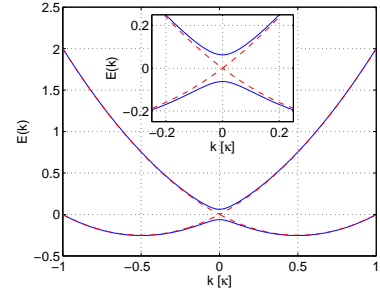,width=.3\textwidth}
   \caption{(Color online). Dispersion relation in the Schr\"{o}dinger limit, with the inset representing the Dirac limit. The red dotted lines correspond to the case of vanishing potential, i.e. $\delta=0$, whereas the blue solid lines correspond to $\delta=mv_D^2/8$. The mixing angle is given by $\theta = 0$.}
   \label{fig:Dispersion_ConstantPotential}
\end{figure}

By applying an additional strong transverse trapping potential to freeze out the transverse degrees of freedom, Eq. (\ref{eq:EffectiveGaugeFieldHamiltonian}) reduces to a 1D equation with
%
%
$2\times 2$-matrices $\textbf{A}=- \kappa\cos\theta\, \hat\sigma_x$, $\textbf{V}=\text{Diag}[V_1, V_1\cos^2\theta + V_3\sin^2\theta]$, and $\boldsymbol\Phi= \kappa^2/2m \text{ Diag}[\sin^2\theta,\sin^2(2\theta)/4]$,
where $\hat\sigma_i$, $i\in\{x,y,z\}$, are the Pauli matrices. Expanding the square in Eq. (\ref{eq:EffectiveGaugeFieldHamiltonian}) and choosing the external potentials as $V_1=\Delta(x)-\kappa^2/2m$, and $V_3=-\Delta(x)-\kappa^2\cos^2\theta/2m$, where $\Delta(x)$ is a detuning we arrive at
\begin{equation}
i\frac{\partial \pmb\Psi}{\partial t} = \left(\frac{\textbf{p}_{x}^{2}}{2m} +  v_D \cos\theta\,\textbf{p}_x\cdot\hat{\sigma}_x + \Delta(x)\hat{\sigma}_z\right)\pmb\Psi,
\label{eq:RandomSchroedingerDirac_unitful}
\end{equation}
with $v_D=\kappa/m$. Eq. (\ref{eq:RandomSchroedingerDirac_unitful}) describes a massive particle with SO coupling moving in a scalar potential $\Delta(x)$. The corresponding dispersion relation for a constant potential $\Delta(x)=\delta$ can easily be calculated and is shown in Fig. (\ref{fig:Dispersion_ConstantPotential})
for the case of $\delta=0$ (red dashed line) and small $\delta\ne 0$ (blue solid line).
By changing the relative intensity of the control lasers one can externally control the value of $\cos\theta$ enabling an easy access to tune the effective SO coupling. In case of a smooth potential and large particle-momenta, {\it i.e.}, $\langle\textbf{p}_{x}^{2}\rangle/2m\gg v_D\langle\vec{p}_x\rangle\cos\theta$, the SO coupling becomes negligible and the problem reduces to two uncoupled massive Schr\"odinger particles moving in an external potential. In the opposite limit, {\it i.e.}, for $\langle\textbf{p}_{x}^{2}\rangle/2m\ll v_D\langle\vec{p}_x\rangle\cos\theta$, Eq. (\ref{eq:RandomSchroedingerDirac_unitful}) reduces to an effective
Dirac equation for a particle with an effective speed of light $c_*=v_D\cos\theta$ and a smooth space-dependent mass $\Delta(x)/c_*^2$ \cite{relbose,zwit}.

By applying Zeeman and Stark shifts, using, e.g., speckle potentials \cite{matterloc1,speckle} or incommensurate optical lattices \cite{matterloc2}, a spatially  fluctuating, random detuning, and thus random potential, can be generated. In the following we assume that the random potential is described by local, Gaussian
white noise,
\begin{equation}
\overline{\Delta(x)\Delta(x')} = \Gamma v_D\delta(x-x'), \quad\overline{\Delta(x)} = 0,
\label{eq:DisorderCorrelationFunctions}
\end{equation}
where the overbar denotes disorder average and all higher order correlation functions factorize. For vanishing SO coupling one expects exponential localization of the particles according to the Anderson scenario \cite{andersontheory,disorder_book,speckle,matterloc1,matterloc2}.
On the other hand, any small SO coupling dominates in the region of small kinetic energy. Neglecting the kinetic energy term proportional to $\langle\vec p_x^2\rangle$ in Eq. (\ref{eq:RandomSchroedingerDirac_unitful}), rather than the SO coupling, one obtains an effective Dirac equation for a particle with a spatially random mass.
This model, also known as the fluctuating gap model (FGM) \cite{Bartosch-1999b,Millis-2000}, is characterized by a Dyson singularity in the density of states (DOS) \cite{Ovchinnikov-1977},
which is a consequence of the chiral symmetry of Eq. (\ref{eq:RandomSchroedingerDirac_unitful}) without kinetic energy leading to a power-law localization.

\begin{figure}[t]
  \epsfig{file=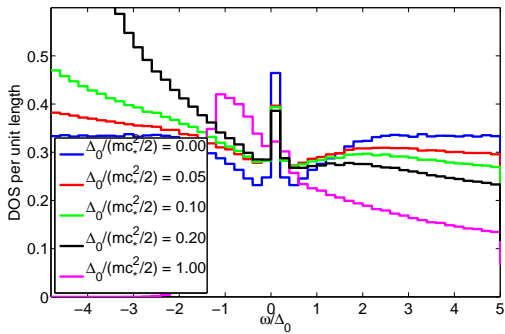,width=.43\textwidth}
   \caption{(Color online). Density of states per unit length for a SO coupled massive Schr\"odinger particle
in a random potential with a finite disorder correlation length \cite{FiniteCorrelationLength} for various values of the band edge $mc_*^2$. As the band edge is moving closer to $\omega=0$ the washing out of the Dyson singularity of the free Dirac case (blue line) becomes more prominent, until the almost purely disordered Schr\"odinger case (magenta line) is reached.}
   \label{fig:DOS_variousJs}
\end{figure}

What determines the crossover from Anderson-like to anomalous localization? In the case of a random potential it is not immediately obvious in which parameter regimes Eq. (\ref{eq:RandomSchroedingerDirac_unitful}) leads to Schr\"odinger or Dirac-like dynamics. To investigate this question we introduce dimensionless units $\xi=\Gamma x /v_D$ and $\tau =\Gamma t$ resulting in
\begin{align}
\hspace*{-.2cm}i\frac{\partial}{\partial \tau}\pmb\Psi=-\frac{\Gamma}{2mv_D^2}\frac{\partial^2}{\partial \xi^2}\pmb\Psi-i\cos\theta\hat{\sigma}_x\frac{\partial}{\partial \xi}\pmb\Psi+\tilde\Delta(\xi)\hat\sigma_z\pmb\Psi,\label{eq:RandomSchroedingerDirac_dimensionless}
\end{align}
where $\tilde\Delta(\xi)=\Delta/\Gamma$. In the following we analyze the DOS in the
system described by Eq. (\ref{eq:RandomSchroedingerDirac_dimensionless}).
For a free Schr\"odinger particle the DOS has a singularity at
$\omega=0$. As can be seen from Fig. \ref{fig:Dispersion_ConstantPotential} the presence of the SO coupling shifts the band-edge of the spectrum away from zero to $\omega_\text{edge}=-m c_*^2/2 = - m v_D^2 \cos^2\theta /2$. In the Dirac limit $mv_D^2 \to \infty$ the band edge moves to infinity.
In the pure Schr\"odinger case weak disorder leads to a smoothing of the band-edge peak \cite{Halperin-1965}, making the DOS analytic, which is associated with the emergence of exponentially localized states (Anderson localization) \cite{Bouchaud-1990}.
To investigate the DOS for our SO-coupled system in the presence of disorder we numerically
simulate Eq. (\ref{eq:RandomSchroedingerDirac_dimensionless}).
As will be the case in any experiment we assume for this a discretized model and furthermore a finite disorder-correlation length \cite{FiniteCorrelationLength}.
(A smaller correlation length is accompanied by slower numerical convergence of the DOS \cite{Millis-2000}.) The mobility edge  in momentum space associated with the finite disorder correlation length is however chosen large enough not to affect the results.
Fig. \ref{fig:DOS_variousJs} shows the DOS for different values of
the dimensionless quantity $\Delta_0/mc_*^2$, where $\Delta_0$ characterizes the rms value
of the Gaussian disorder amplitude in the discretized model. It is related to the continuum quantity via $\Gamma=\Delta_0^2/(v_D n_\text{kinks})$, with $n_\text{kinks}$ being the impurity density.
%
%
For $mc_*^2 \to \infty$, i.e. in the Dirac limit, one recognizes a Dyson-like singularity at $\omega=0$.
The emergence of a Dyson singularity can in general be taken as an indicator for anomalous, i.e. non-exponential, localization properties \cite{disorder_book}. In the random-mass Dirac model or FGM the Dyson singularity has been shown to lead to power-law correlations \cite{powerlaw}.
As $mc_*^2\sim m v_D^2\cos^2\theta$ decreases the band-edge approaches the singularity. As
a consequence the singularity is smoothend out. It should be noted at this point that a true
singularity is present only in the exact Dirac limit, but as can be seen
from Fig. (\ref{fig:DOS_variousJs}) a pronounced peak survives as long as the SO coupling, i.e. $mc_*^2$, is large.
Thus one expects in this case a localization scenario
where power-law correlations dominate for very long time scales. To quantify this we define
 following \cite{Millis-2000} the width $\Delta\omega_D$ of the Dyson singularity as the minimum of the DOS,
yielding $\Delta\omega_D = \alpha \Gamma$, where $\alpha=0.6257...$. Hence, to obtain dynamics associated with the Dyson singularity the condition
\begin{equation}
 mv_D^2 \cos^2\theta \gg \Gamma,
 \label{eq:ConditionDysonSingularityDynamics}
\end{equation}
should be satisfied in order to ensure that the band-edge is sufficiently far away from the singularity.

\begin{figure}[t!]
\epsfig{file=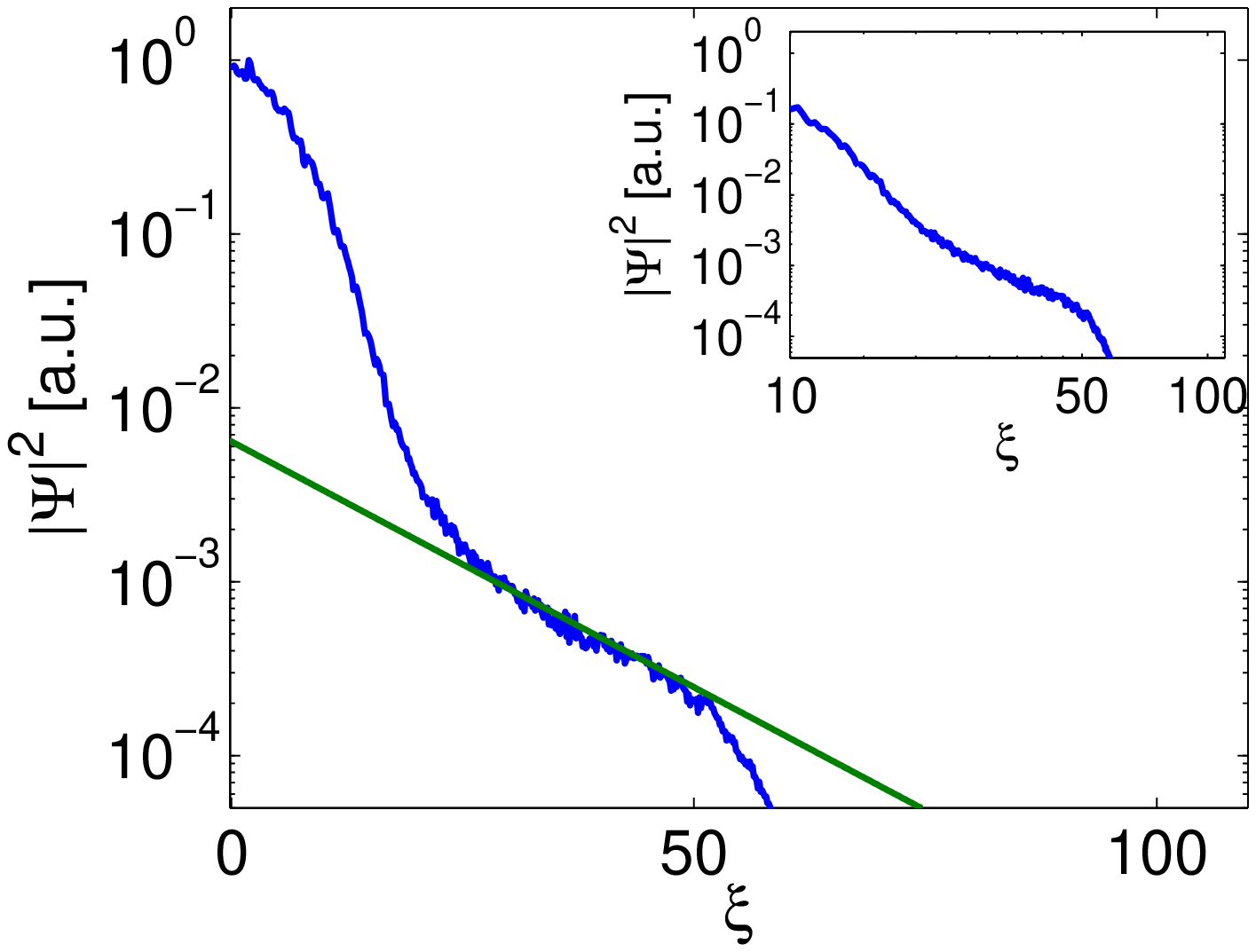,width=.24\textwidth}
\hspace*{-.5cm}
\epsfig{file=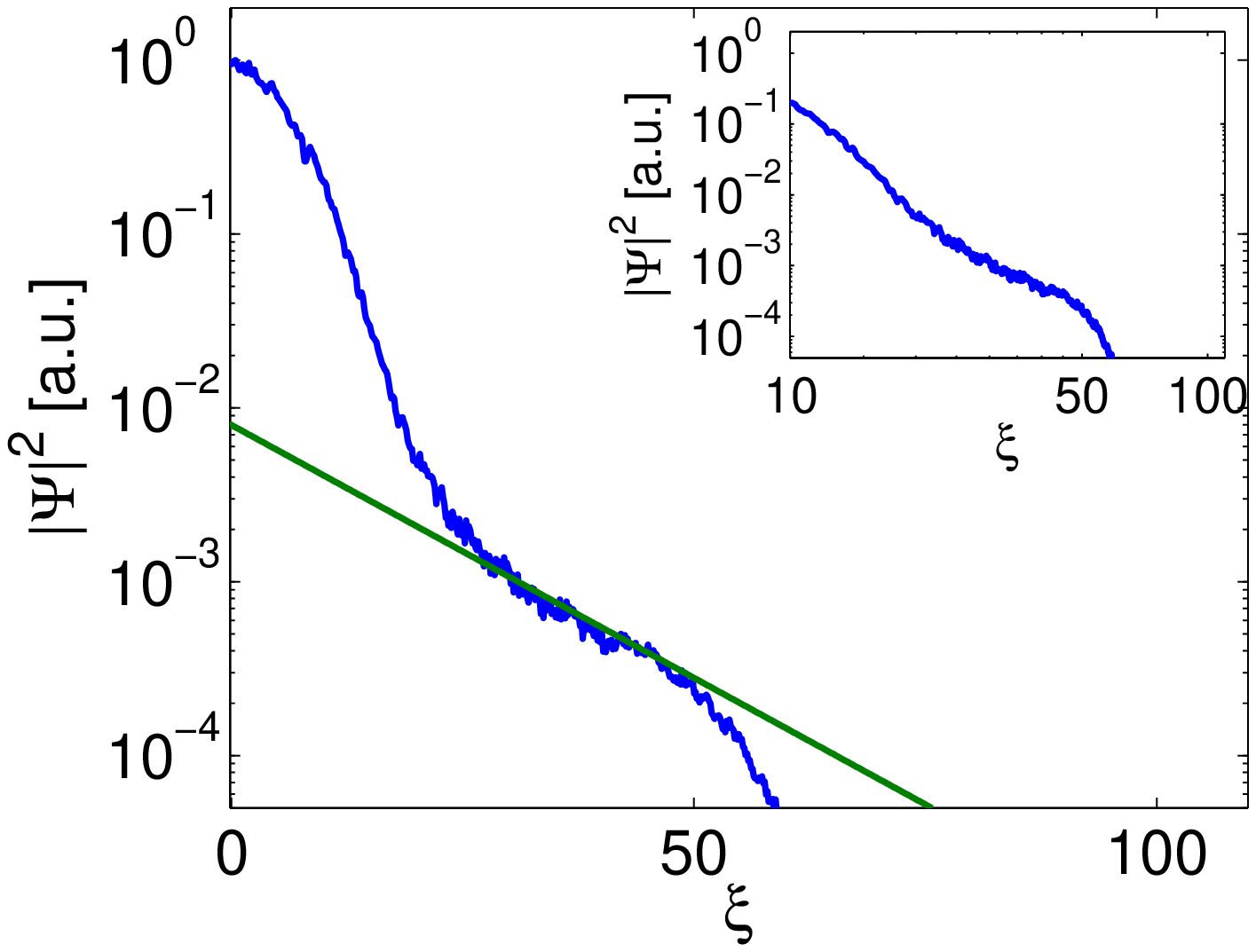,width=.24\textwidth}
\epsfig{file=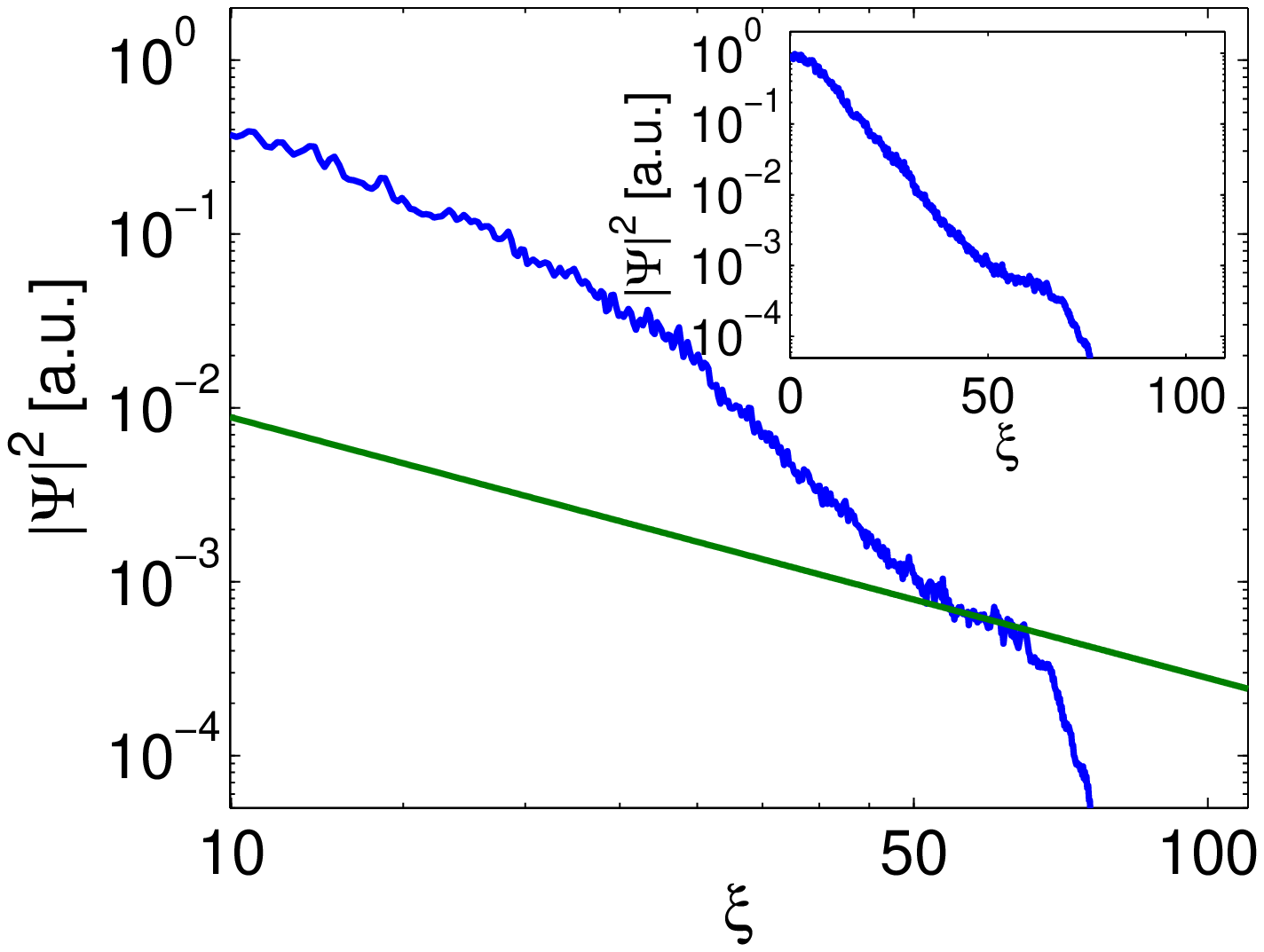,width=.24\textwidth}
\hspace*{-.5cm}
\epsfig{file=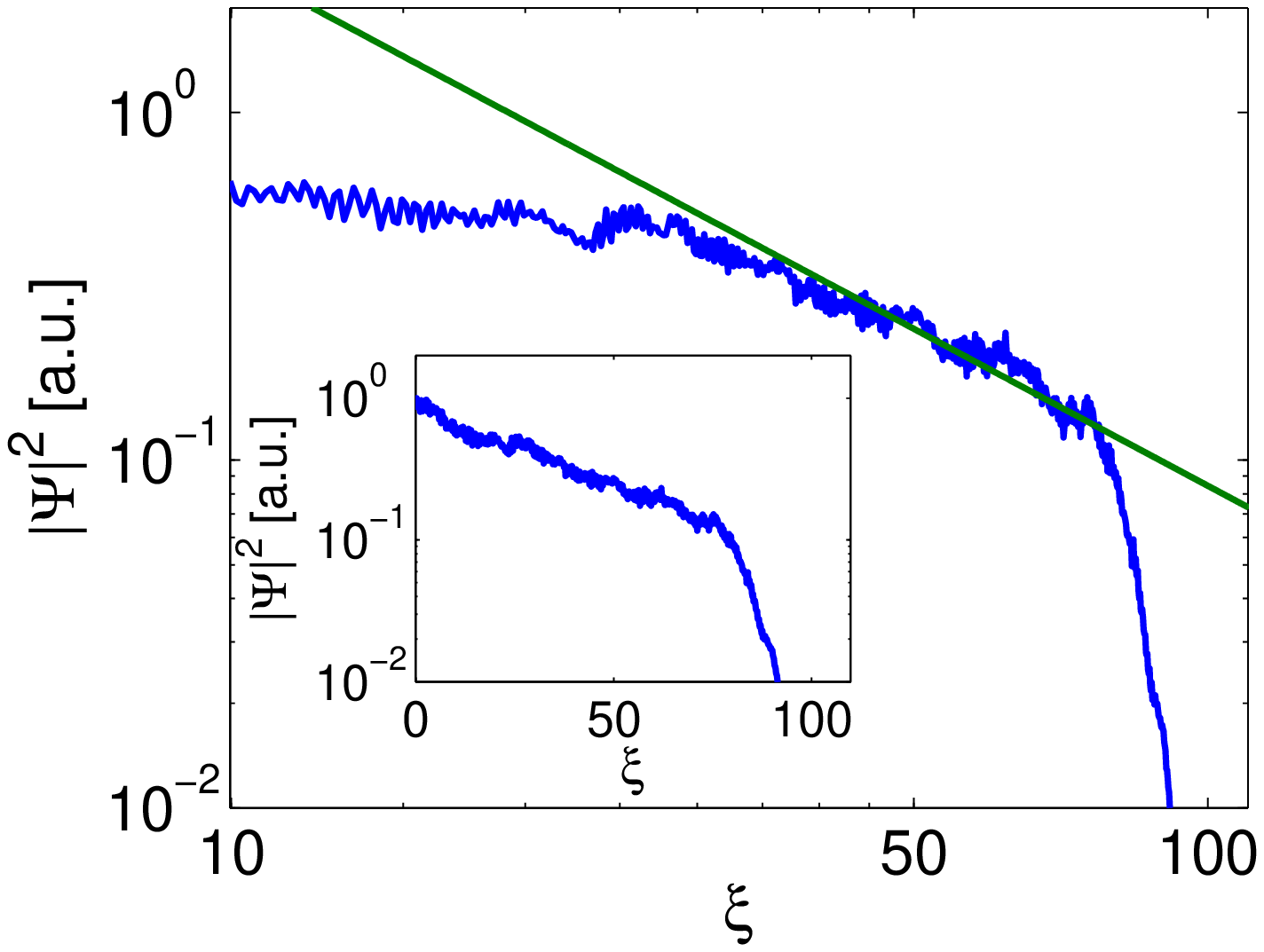,width=.24\textwidth}
   \caption{(Color online). Stationary density profiles of an initially localized wave packet of size $L_0 \Gamma/v_D=7.5$ after a time evolution $\Gamma t=80$ according to eq. (\ref{eq:RandomSchroedingerDirac_dimensionless}) with $\Gamma/mv_D^2=0.1$ after averaging over 100 independent realizations. The system size is $2L\Gamma/v_D=240$ and $n_\text{kinks}=6.25\Gamma/v_D$, i.e. $\Delta_0^2/\Gamma^2=6.25$. (\textit{Top left, right}) Schr\"odinger limit with $\cos\theta=(0.01,0.05)$. The green (straight) line corresponds to an exponential fit showing the expected Anderson localization. The inset shows the same density in log-log. (\textit{Bottom left}) Density in the crossover regime with $\cos\theta=0.5$ in log-log representation (semi-log in the inset) and a power-law fit (green straight line) with exponent $3/2$. (\textit{Bottom right}) Dirac limit with $\cos\theta=1$ in log-log; the power-law behavior is clearly visible for large distances $\xi\gg 1$.}
   \label{fig:Density_Profiles}
\end{figure}

The emergence of the Dyson singularity affects the dynamics of the system around $\omega=0$. Experimentally this is reflected in a drastically different
behavior of the density profile after a long-time expansion of an initially localized wave-packet. From Eq.(\ref{eq:ConditionDysonSingularityDynamics})
we expect that for $\cos^2\theta\gg\Gamma/mv_D^2$ the effective Dirac dynamics along with the creation of anomalous power-law correlations
dominates the system whereas in the opposite limit we expect Anderson-like localization. The power-law behavior of correlations will also be reflected
in the density distribution of an expanding wavepacket, which is experimentally much easier accessible. To see this we note, that in the power-law case there is no
intrinsic length scale. Thus we may assume that an initially well localized wave-packet will show the same behavior as a wavepacket with an initial
delta-distribution $\pmb\Psi(x,t=0)=\delta(x) \pmb{\chi}$ with $\pmb\chi$ being a constant vector with unity length. The time-evolution of this wave-packet is
given by
 $\pmb\Psi(x,t)\,=\,\sum\limits_n \left[\pmb\phi_n^*(0)\cdot\pmb\chi \right]\pmb\phi_n(x)\e^{-i\omega_nt}$,
where $\pmb\phi_n(x)$ is the stationary state of a particular disorder realization belonging to the state of energy $\omega_n$.
It immediately follows that the long-time evolution is given by
 $\overline{|\pmb\Psi(x,t\rightarrow\infty)|^2}=\sum\limits_n\overline{|\pmb\phi_n(0)|^2|\pmb\phi_n(x)|^2}$.
The right hand side of this expression corresponds to the localization criterion defined in \cite{disorder_book}, and it is thus sufficient to investigate the long-time behavior of the density in order to determine localization properties.
In the Schr\"odinger limit all correlations of the type $C_n(x)=\overline{|\pmb\phi_n(0)|^2|\pmb\phi_n(x)|^2}$ decay exponentially \cite{disorder_book}, i.e. $C_n(x)\sim\e^{-|x|/L_\text{loc}}$, where the localization length $L_\text{loc}$ is at most of the order of the system size. In the FGM the typical
localization length of correlation functions scales as $L_\text{loc}\sim\left|\text{ln}\epsilon\right|^2$ \cite{randdirac}, where $\epsilon$ is the distance
from the band-center. For small energies this length is much larger than the system size and consequently the correlation functions should be similar
to the zero-energy correlation function decaying with a power law scaling as $C_n(x)\sim \left(\Gamma |x|/v_D\right)^{-3/2}$ at large distances \cite{randdirac}. Accordingly one expects quite different scaling behavior in the two different limits of large and small SO coupling. To confirm the predictions we
numerically simulated the time-evolution according to Eq. (\ref{eq:RandomSchroedingerDirac_unitful}) of an initially localized wave-packet of the form $\pmb\Psi(x,t=0)= \left(2\sqrt{2\pi}L_0 \text{erf}\left(L/\sqrt{2}L_0 \right) \right)^{-1/2}\exp\left(-x^2/4L_0^2 \right)(1,i)^T$ of width $L_0$ in a system of total length $2L$. The results, shown in Fig. (\ref{fig:Density_Profiles}), depict the density profile for different SO coupling strengths and $\Gamma/mv_D^2=0.1$ after a sufficiently long time evolution $T$.
%
%
For $\cos\theta=(0.01,0.05)$ one can clearly see the exponentially localized wings of the density reminiscent of Anderson localization in cold atom systems \cite{matterloc1,matterloc2,andersontheory}. Increasing the SO coupling strength leads to an intermediate regime at $\cos\theta=0.5$ where neither exponential nor power-law localization can be determined. Finally for strong SO coupling, i.e. $\cos\theta=1$ a fit with a power-law reveals that the density correlation scales as the expected $3/2$ power-law at large distances in agreement with the predictions of the FGM.

To understand the $3/2$ power-law exponent and show its connection to the formation of a zero-energy state, we perform a gauge transformation of Eq. (\ref{eq:RandomSchroedingerDirac_unitful}) with $\pmb\Psi=\exp\left(-imc_*x\hat\sigma_x\right)\pmb\Phi$
%
%
and subsequently substitute the Ansatz $\pmb\Phi=\pmb\chi_+\exp\left(i mc_*x\right)+\pmb\chi_-\exp\left(-imc_*x\right)$ giving (details see \cite{Supplementary1})
\begin{align}
\hspace*{-.5cm} i\frac{\partial\pmb\chi}{\partial t}=-i c_* \hat\tau_z\otimes\mathds{1}\frac{\partial\pmb\chi}{\partial
x} +\frac{\Delta(x)}{2}\left(\hat\tau_x\otimes\hat\sigma_z+\hat\tau_y\otimes\hat\sigma_y \right)\pmb\chi,
 \label{eq:RandomMassDiracEquation}
\end{align}
where terms proportional to $\partial_x^2\pmb\chi$ or fast oscillating exponentials were dropped under the assumption $\Gamma/mv_D^2 \ll \cos^2\theta$. Further we defined the four-component object $\pmb\chi=(\pmb\chi_+,\pmb\chi_-)^T$ and $\hat\sigma_i$, $\hat\tau_i$, $i\in\{x,y,z\}$, act on the momentum or internal degree of freedom, respectively. Equation (\ref{eq:RandomMassDiracEquation}) is a generalization of the model considered in \cite{powerlaw,randdirac}. It possesses a zero-energy (mid-gap) state given by
%
$\pmb\chi(x)\sim\exp\left\{\pm \hat\alpha/2c_* \int\limits_{-\infty}^x\text{d}y \Delta(y) \right\}\pmb\chi_0$.
%
The matrix $\hat\alpha=\left(\hat\tau_y\otimes\hat\sigma_z+\hat\tau_x\otimes\hat\sigma_y\right)$ has eigenvalues $(1,-1,0,0)$. Choosing $\pmb\chi_0$ to be an eigenvector with eigenvalue $\pm 1$ we obtain a power-law intensity correlation scaling as $C(x)\sim (\Gamma |x|/v_D)^{-3/2}$ \cite{powerlaw,randdirac} in full agreement with the numerical simulations.

In summary, we have investigated the one-dimensional dynamics of a SO-coupled massive Schr\"odinger particle subject to a $\delta$-correlated disorder potential. For weak SO coupling the system is equivalent to two independent Schr\"odinger particles with diagonal disorder. In the opposite limit the system is described by the random-mass Dirac model with off-diagonal disorder. The model can be implemented with current state-of-the-art techniques by using atomic dark-states in cold atom systems. We showed by calculating the systems' DOS and direct numerical simulation of the time evolution of an expanding wave packet that there is a crossover from an exponential (Anderson) localized regime to a power-law regime governed by a Dyson singularity when varying the strength of the SO coupling relative to the disorder strength. The crossover can be observed by expansion of an initially localized wavepacket with an appropriately chosen width of the momentum distribution in a random potential.

Future investigations should include effects of interactions, which will allow for studying models such as the relativistic Thirring model \cite{Thirring-1958} for fermions as well as for bosons with a random mass. Interactions may well accelerate the crossover to non-exponential behavior as they tend  to delocalize particles.

\textit{Acknowledgements.--} We wish to thank B. Halperin, C. Hooley, M. Merkl and D. Muth for useful discussions. We acknowledge support from the EPSRC Scottish Doctoral Training Centre in Condensed Matter Physics (M.J.E.), the DFG through the project UN 280/1 (R.G.U.), the SFB TR49 (M.F.) and the Harvard Quantum Optics Center (J.O.).


\begin{thebibliography}{99}
\bibitem{FiniteCorrelationLength} For the numerics we used an exponentially correlated disorder according to $\overline{\Delta_i\Delta_j}=\Delta_0^2\Delta x/L_\text{corr}\exp\{-|i-j|\Delta x/2L_\text{corr}\}$ with correlation length $L_\text{corr} = 2 \Delta x$, where $\Delta x$ is the discretization length.
\end{thebibliography}
\end{document}